\documentclass[notitlepage, nofootinbib,12pt]{revtex4-1}

\newcommand{\ket}[1]{{|{#1}\rangle}}
\newcommand{\ip}[2]{{\langle{#1}|{#2}\rangle}}
\usepackage{amssymb,amsmath}
\usepackage{epsfig}
\usepackage{epstopdf}
\usepackage{bm}

\begin{document}

\title{Action Duality: A Constructive Principle\\ for Quantum Foundations}

\author{K.B. Wharton$^{1,2}$, D.J. Miller$^2$ and Huw Price$^2$}
\affiliation{$^1$Department of Physics and Astronomy, San Jos\'e State University,\\ San Jos\'e, CA USA\\ $^2$Centre for Time, University of Sydney, Sydney, NSW Australia}


\begin{abstract}
An analysis of the path-integral approach to quantum theory motivates the hypothesis that two experiments with the same classical action should have dual ontological descriptions.  If correct, this hypothesis would not only constrain realistic interpretations of quantum theory, but would also act as a constructive principle, allowing any realistic model of one experiment to generate a corresponding model for its action-dual.  Two pairs of action-dual experiments are presented, including one experiment that violates the Bell inequality and yet is action-dual to a single particle.  The implications generally support retrodictive and retrocausal interpretations.

\end{abstract}

\maketitle

\section{Introduction}

The success of the Feynman path-integral, especially in its extension to quantum field theory (QFT), makes it natural to wonder whether  the path-integral framework embodies some important truth about the nature of the quantum world.  However, the framework does not appear to have a natural realistic interpretation. In particular, it is widely accepted, as it was by Feynman himself, that it cannot be interpreted as implying that a quantum system actually follows one of the trajectories over which the path-integral is computed.  (The problem is that path-integral sums complex amplitudes, not classical probabilities. See \cite{Feynman, FeynmanHibbs, Schulman} for further discussion of this point.)  

Still, there may be less direct options for allowing the path-integral framework to guide the development of realistic models of the quantum world. For present purposes, we define a realistic model, intuitively, as one which attempts to answer the question ``What is happening between measurements?''  We shall not defend or even assume the claim that the construction of such models is preferable to rival approaches in quantum foundations. Our aim is simply to propose what we take to be a new way of relating the path-integral framework to the project of constructing realistic models.

In outline, our proposal is based on the hypothesis that the \emph{symmetries} of the path-integral should be mirrored in the symmetries of any ontology underlying quantum theory.  We shall call this the \emph{Feynman Integral Symmetry Hypothesis} (FISH), to be precisely defined in Section II.  As we shall show, FISH's effect is not only to \emph{constrain} potential ontologies, by excluding models whose ontologies do not share the symmetries of the path-integral; but also to allow us to \emph{construct} ontological models of specific experiments, when we already have a model of some other experiment, to which the given experiment bears an appropriate path-integral symmetry.  (It is true that some of the models constructed in this way are counterintuitive, by ordinary standards. However, the suggestion we are exploring is that symmetries of the path-integral might be a better guide to the nature of the quantum world than are the intuitions these models seem to offend.)

FISH is closely related to the ``action symmetry'', used in a recent challenge to the widely accepted view that quantum mechanics reveals the presence of action-at-a-distance in the quantum world.\cite{EPW}  By comparing one-photon and two-photon experiments with the same path-integral mathematics, the authors of that paper present the orthodox view with a dilemma: either (i) action-at-a-distance is much more widespread than it is usually thought to be, or (ii) action-at-a-distance is much less widespread than it is usually thought to be.  As they note, their opponent can avoid this dilemma by rejecting the action symmetry (\emph{i.e.}, in our present terminology, by rejecting FISH). But they conclude that in the absence of any strong independent reason for rejecting this symmetry,  the case for action-at-a-distance is considerably weaker than it is usually thought to be.

The present paper aims to extend these results, by providing further illustrations of the ubiquity of the symmetries, or ``dualities'', implied by the path-integral framework. Section II describes the central action duality, on which FISH is based. Sections III and IV then present two applications of this duality.  Each involves a pair of experiments in which it is possible to calculate the path-integral explicitly, and in which every step of the mathematics is identical in both experiments.  Given these symmetries, FISH then leads to particular conclusions about how these experiments may be interpreted, within a realistic framework.  Finally, Section V generalizes these results in a manner similar to recent work by Leifer \cite{Leifer}, which was in turn based on an isomorphism noted by Jamiolkowski \cite{Jam}.  

Both in classical and quantum theory, least-action and path-integral principles seem to be generally regarded as useful mathematical tools with few physical implications.  FISH amounts to the proposal that realist approaches to quantum foundations should take this mathematical framework more seriously.  It proposes that if we are interested in finding realistic foundations for quantum phenomena, we should take on trust the symmetries of the path-integral and see where this policy takes us, even when it conflicts with naive intuitions.

\section{The Action Duality}

It is a remarkable fact that quantum probabilities can be expressed in terms of the classical action $S$, in both non-relativistic quantum mechanics (QM) and in quantum field theory (QFT).  In the case of QM, the Feynman path-integral \cite{Feynman, FeynmanHibbs} expresses the unnormalized joint probability of any initial wavefunction $\psi_i(x)$ measured at time $t=t_i$ followed with any final wavefunction $\psi_f(x)$ measured at time $t=t_f$ as
\begin{equation}
\label{eq:fpi}
P(\psi_i,\psi_f)=\left| \int\int \left(\sum_{x,t_i\to y,t_f} e^{iS/\hbar}\right) \psi_f^*(y) \psi_i(x) \, dx \, dy \right| ^2.
\end{equation}
Here the classical action is $S=\int L \, dt$, evaluated by integrating the classical particle Lagrangian $L$ along a given path specified in the so-called ``sum over all paths".  The above expression (\ref{eq:fpi}) can be trivially extended to multiple spatial dimensions or multiple particles simply by extending the integration to cover all of the parameters in $\psi$, as well as generalizing the sum to cover the paths of all the particles.  It is crucial to note that the only portion of (\ref{eq:fpi}) that mathematically encodes any events between times $t_i$ and $t_f$ is the action $S$ itself.  

While symmetric between past and future, joint probabilities are not as useful as conditional probabilities, at least when it comes to making predictions.  For this reason, one typically interprets (\ref{eq:fpi}) directly as a conditional probability $P(\psi_f|\psi_i)$, which can be done with the appropriate normalization constant.  But there is another, more careful route to this same answer that does not require any explicit distinction between past and future.  Given a well-defined preparation at $t\!=\!t_i$, any analysis of (\ref{eq:fpi}) need only consider some particular initial wavefunction $\psi_i$ rather than the larger space corresponding to all possible initial measurements.  Using the particular type of measurement (position, momentum, etc.) made at $t=t_f$, quantum measurement theory defines a space of allowed wavefunctions $\psi_n$ that are eigenstates of the corresponding measurement operator.  The standard probability rule
\begin{equation}
\label{eq:cp}
P(\psi_f|\psi_i) = \frac {P(\psi_i,\psi_f)}{\sum_n P(\psi_i,\psi_n)}
\end{equation}
then generates the conditional probabilities.  Here the subscript $f$ is performing double duty by also specifying the particular value of $n$ for which one wants to calculate the conditional probability.  Another advantage to this approach is that it forces an automatic normalization of the resulting conditional probabilities, such that no normalization coefficient is needed in (\ref{eq:fpi}).

Alternatively -- though much less usefully, in practice -- one can use the same joint probabilities to generate conditional probabilities for possible initial wavefunctions $\psi_i$ given some certain final wavefunction $\psi_f$.  (Such a ``reversed'' conditional probability $P(\psi_i|\psi_f)$ would only be correct when $\psi_i$ is unbiased by events in its past, rather than non-randomly preselected by an experimenter.)  The fundamental use of joint probability in (\ref{eq:fpi}) makes it clearer that path-integrals are a fully time-symmetric formalism, and the construction of conditional probabilities based on either pre- or post-selection is a choice, not an asymmetry.  To make this even clearer, note that time-reversal not only switches the ``$i$'' and ``$f$'' subscripts, but also complex conjugates $\psi_i$ and $\psi_f$ such that time-odd quantities like momentum change sign.  The net result leaves (\ref{eq:fpi}) unaffected under time-reversal so long as the scalar action (and underlying Lagrangian) are invariant under such a transformation.  

Extending the path-integral formalism to QFT leads to an even simpler expression for the joint probability (although much more difficult to evaluate).  The initial and final field states $\phi_i(x)$ and $\phi_f(x)$ now merely act as constraints on the possible field configurations spanned by a functional integral:
\begin{equation}
\label{eq:qft}
P(\phi_i,\phi_f)=\left| \int {\cal D}\phi \, e^{iS[\phi]/\hbar} \right| ^2.
\end{equation}
Here the classical action is defined via $S[\phi]=\int {\cal L}(\phi,\partial_\mu \phi)\, d^4x$, integrating the classical Lagrangian density ${\cal L}$ (specified by the field and its local derivatives) over the spacetime volume bounded by the hypersurfaces on which $\phi_i$ and $\phi_f$ are defined/measured.  The functional integral in (\ref{eq:qft}) is over all field configurations consistent with the initial and final field values; as before, multiple fields can be easily accommodated by extending this integral.  Conditional probabilities can again be generated via the procedure (\ref{eq:cp}).

When calculating probabilities in QFT, it continues to be the case that the classical action $S$ is the only mathematical object that in any way encodes events between the two measurements.  This has profound consequences, in that the mathematics implies a duality between any two experiments with the same classical action -- the two experiments will always have the same joint probabilities.  As the action is just a spacetime integral of the classical Lagrangian density, the implication is a deep connection between ${\cal L}$ and quantum phenomena.  While the mathematics does not provide us with the nature of this connection, symmetry considerations motivate the following hypothesis:
\begin{quote}
{\bf Feynman Integral Symmetry Hypothesis (FISH):} For any two experiments with an action duality (a well-defined spacetime transformation that maps the classical Lagrangian density of one experiment onto the classical Lagrangian density of the other), any realistic ontology must also map between the two experiments under the same spacetime transformation.
\end{quote}
One way to express this symmetry-based motivation is to note that if FISH were false, the observable probabilities of action-dual experiments would contain a symmetry that was not respected by the ontology.  In other words, the failure of FISH would imply ``asymmetries [in the ontology] which do not appear to be inherent in the phenomena'', to quote the most widely-known example of using a mathematical duality to inform physical significance \cite{Einstein}.

The fact that action-dual experiments have the same joint probabilities has long been used as a calculational tool for relating certain scattering amplitudes -- the so-called ``crossing symmetry'' in QFT -- but until recently \cite{EPW} this fact had not been utilized in the context of interpreting quantum theory itself.  If FISH holds, action dualities would correspond to dualities in the underlying ontology; this in turn would become a new constructive principle for quantum foundations.  In other words, given any realistic interpretation of one experiment, the dual interpretation of any action-dual experiment can be constructed via a well-defined spacetime transformation.  While the ontologies constructed in this manner are only as valid as the original interpretation, this procedure nevertheless constructs well-defined ontologies that might have otherwise not been considered.  

For the purposes of calculating the probabilities of experimental outcomes, the path-in tegral formulation of QM is equivalent to the other representations: the Schr\"odinger representation, the Heisenberg representation, etc.  But it does not follow that each representation has the same ontological implications.  There are several reasons why the use of the classical action in the path-integral makes it particularly well suited to generating promising models of what is happening {\em between} measurements.  First, the action is an integral over spacetime, not configuration space, regardless of how many particles and/or fields are being considered.  This makes it more likely that any implied ontology will also reside in spacetime, arguably one of the crucial requirements for it to be considered ``realistic''.  Also, the ability to start with a local and Lorentz-invariant Lagrangian density would presumably make it more likely that the implied ontologies would incorporate these same important features -- certainly more likely than starting with the non-relativistic Schr\"odinger equation.

As a further advantage, the action $S$ is explicitly {\em classical}, only incorporating known fields/particles.  This certainly does not rule out realistic models that postulate non-classical behavior of those classical entities, or even new fundamental fields and/or particles.  But FISH strongly constrains any such non-classical elements, requiring that they not break the classical action duality between any two experiments.  FISH seems quite reasonable in this context: Assuming that the probabilities generated from QFT are correct, those two experiments (having the same classical action) would continue to have the same joint probabilities.  But if some non-classical physics were to break this duality, then these probabilities would now presumably derive from different underlying reasons.  Any such ``broken-duality'' model would be in the curious position of predicting experimental correlations with a symmetry reflected by the obsolete classical action but not by the newly proposed ontology. 

The previously published application of the action duality \cite{EPW} considered the following two experiments.  (A) One photon, sent through a polarizer set at angle $\alpha$, reflects at a mirror, and then passes through a polarizer set at angle $\beta$.  (B) Two entangled photons produced at some source with identical (but unknown) polarizations; one is sent through a polarizer set at $\alpha$, the other is sent through a polarizer set at $\beta$.  There is an action duality between these two experiments; an action-preserving spacetime transformation that relates each possible pair of inputs/outputs in (A) with each possible pair of outputs in (B).  As a direct result, the joint probabilities in (A) and (B) are exactly the same (both violate the Bell inequality).

In this example, the power of the action duality can be easily demonstrated.   If one postulated some new faster-than-light communication channel between the two polarizers in (B), but no corresponding new channel to connect the time-like separated polarizers in (A), the classical action duality would be broken.  The fact that the correlations in (A) and (B) continued to be identical, and continued to be correctly described by an equation (the {\it same} equation) with no terms corresponding to this new channel, would be a profound mystery -- again, the new theory would have an ``asymmetry not inherent in the phenomena".  Additionally, if some mathematical structure that resided in configuration space rather than spacetime was invoked as an explanation of the correlations in (B) but not (A), the action duality would still be broken if there were any effects of this structure on anything that {\it did} reside in spacetime.

In the following two sections, two further examples will be developed that highlight the capabilities of the action duality -- not merely as a constraint, but as a new constructive principle.  Section III highlights the fact that any literal interpretation of a traditional single-particle QM wavefunction breaks the action duality.  The example in section IV explores how this duality can be applied to situations with Bell inequality violations, utilizing an experimental geometry which allows a more explicit analysis than is possible in the previous publication \cite{EPW}.

\section{Action Duality under Time Reversal}

As a detailed example of the capabilities of the action duality, consider Experiment A1 shown in Figure 1.  Here a single photon at $C$ is injected into a double Mach-Zehnder interferometer (the solid rectangles are mirrors, the empty rectangles are ideal 50/50 beamsplitters). Detectors are placed at the outputs $A$ and $B$; only one will fire for a single photon input.  The circle $E$ is an element that introduces a phase delay of $\alpha$ into one of the paths.  The phase delay on reflection by the 50/50 beamsplitters is $\pi/2$, but any phase delays induced by the mirrors or distances are the same for all paths and will be omitted for simplicity (as is usual).

\begin{figure}[htb]
\centerline{\includegraphics[width=.15\textwidth]{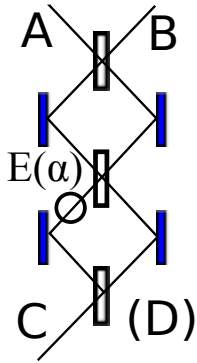}}
\caption{Experiment A1: a double Mach-Zehnder interferometer for a single photon input at $C$.  The circle represents a phase delay element $E$ set to a phase delay angle $\alpha$.}
\label{Figure:a1}
\end{figure}

It is not immediately obvious that the probabilities of the detectors $A$ and $B$ firing are always 50\%, independent of the value of $\alpha$.  Even an explicit calculation of this fact sheds little light on exactly why $E$ doesn't have any impact on the outcome (as it would if the central beamsplitter were removed).  But using the fact that for a 50/50 beamsplitter, $\ket{input}=(\ket{T}+i\ket{R})/\sqrt{2}$ (where $\ket{T}$ is the transmitted wavefunction and $\ket{R}$ is the reflected wavefunction), standard quantum theory reveals how this geometry transforms the initial wavefunction. We can label the left and right paths in Fig.~\ref{Figure:a1} as $L1$, $R1$ for the bottom interferometer stage and $L2$, $R2$ for the top interferometer stage. Then
\begin{align}
\label{eq:eq1}
\ket{C} & =\frac{1}{\sqrt{2}}(i\ket{L1} + \ket{R1}) \notag \\
& = \frac{1}{2}[(1-e^{i\alpha})\ket{L2}+i(1+e^{i\alpha}) \ket{R2}] \notag \\ 
& = \frac{1}{\sqrt{2}}(i\ket{A}-e^{i\alpha}\ket{B}),
\end{align}
and the Born rule probability $P(A|C)=|\ip{A}{C}|^2=1/2$ is indeed independent of $\alpha$.

This same result can be easily extracted from the path-integral formulation.  For a final measurement of $A$, the joint probability $P(C,A)$ can be written in the form
\begin{equation}
\label{eq:fpi3}
P(C,A) = \left| \sum_{A\to C} e^{iS/\hbar} \right|^2.
\end{equation}
This is simpler than (\ref{eq:fpi}) because we have highly localized initial and final conditions.  Technically, this sum should be over an infinitely dense set of possible paths, but as in the ``sum over histories'' literature \cite{Sorkin}, one can ``coarse-grain'' these into classical paths with different phases.  In other words, one need not consider all of the infinitesimally different ways that a path could reflect off of a mirror, because applying (\ref{eq:fpi3}) to those paths would just sum to the ray-tracing limit of classical optics, equivalent to a single term where $S/\hbar$ is equal to the accumulated phase $\theta$ along the classical path.  Furthermore, because all of the beamsplitters in this example are 50/50, each coarse-grained path has the same magnitude, and all of the physics is simply encoded in the relative phases of these classical paths.

To get from $C$ to $A$ in experiment A1 there are only 4 such coarse-grained paths; a choice of left or right at each of the first two beamsplitters.  Only relative phase differences will matter in this calculation, these being a $\pi/2$ phase for each reflection off of a 50/50 beamsplitter and an $\alpha$ phase for a pass through $E$.    So, for example, one has $\theta_{L1R2}=\alpha +\pi/2$, using the notation ``L1R2'' to refer to the path that veers left at the first beamsplitter and then right at the second beamsplitter, on its way to $A$.  All together, one has the unnormalized joint probability
\begin{align}
\label{eq:sumA1}
P(C,A) & =  \left| e^{i\theta_{L1L2}}+ e^{i\theta_{L1R2}}+ e^{i\theta_{R1L2}}+ e^{i\theta_{R1R2}} \right|^2 \\
 &=  \left| e^{i(\alpha+3\pi/2)}+e^{i(\alpha+\pi/2)} + e^{i\pi/2}+e^{i\pi/2} \right|^2 =4 \notag
\end{align}
which is independent of $\alpha$.  The same calculation for $P(C,B)$ yields
\begin{align}
\label{eq:sumA2}
P(C,B) & =  \left| e^{i\theta_{L1L2}}+ e^{i\theta_{L1R2}}+ e^{i\theta_{R1L2}}+ e^{i\theta_{R1R2}} \right|^2 \\
 &=  \left| e^{i(\alpha+\pi)}+e^{i(\alpha+\pi)} + e^{0}+e^{i\pi} \right|^2 =4, \notag
\end{align}
which is also independent of $\alpha$.  The conditional probability is then simply $P(A|C)= 4/(4+4)=50\%$.

While yielding the same conclusion as the standard QM calculation, this alternate mathematics offers some new insight as to why $\alpha$ might be irrelevant to the end result.  Specifically, $\alpha$ vanishes from $P(C,A)$ because the first two terms in (\ref{eq:sumA1}) cancel.  These two terms are the only terms which contain the ``L1'' portion of the path which passes through $E$.  Therefore, restricting the path-integral to only  paths which pass through $E$ yields a zero ``partial amplitude''.  From this one might be tempted to conclude that nothing ontological passes through $E$, as this would naturally explain the irrelevance of $\alpha$ to the outcome.  Although such a conclusion would only apply to an $A$-outcome, the corresponding calculation of $P(C,B)$ shows that there is a zero ``partial amplitude'' for paths that include ``R1'' (because the last two terms in (\ref{eq:sumA2}) cancel).  The corresponding implication would be that nothing {\it fails} to pass through $E$ in the case of a $B$-outcome, making $\alpha$ an irrelevant global phase.

But although the path-integral mathematics offers these new explanatory features, it is unclear that one can take them too seriously.  The main problem with the analysis in the previous paragraph is that it stands or falls with a well-defined interpretation of the path-integral mathematics itself, and attempts at such an interpretation come with significant problems as noted in the opening paragraph.  And yet throwing out any analysis of this mathematics entirely seems too drastic.  With that in mind, FISH provides a way to usefully analyze the path-integral mathematics without committing to a particular interpretation, as can be seen by setting up an action-dual experiment.

Experiment A2, as shown in Figure 2, is simply the time-reverse of A1.  Here the input is a single photon at $A$, and the detectors are at $C$ and $D$.  The action duality between A1 and A2 follows from the time-symmetry of the classical action, although one must be careful to compare the appropriate boundary conditions.  For example, a detection at $D$ in A2 is not action-dual to a detection at $B$ in A1, but instead would be dual to a photon sent into A1's empty channel (on the lower right of Figure 1) and then detected at $A$.

\begin{figure}[htb]
\centerline{\includegraphics[width=.15\textwidth]{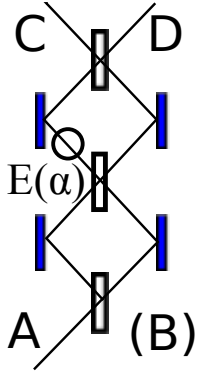}}
\caption{Experiment A2: a double Mach-Zehnder interferometer for a single photon input at $A$.  The circle represents a phase delay element $E$ set to a phase delay angle $\alpha$.}
\label{Figure:a2}
\end{figure}

In experiment A2, a standard QM analysis quickly shows that the first interferometer entirely cancels the amplitude of the component of the wavefunction that passes through $E$; the quantity $\alpha$ is therefore ``intuitively'' not important, and the standard calculation indeed yields
\begin{equation}
\label{eq:eq2}
\ket{A}=\frac{i}{\sqrt{2}}\ket{C}-\frac{1}{\sqrt{2}}\ket{D}.
\end{equation}
Note that this intermediate step to calculating the probability $|\ip{C}{A}|^2$ differs from the corresponding step (\ref{eq:eq1}) in A1, despite the fact that the result is identical.  In the standard QM framework, then, it is hard to see how the ``intuitive'' explanation of $\alpha$-independence in A2 can be mapped to the same result for A1.

The path-integral mathematics, on the other hand, is explicitly identical between these two experiments; not only is $P(A,C)$ the same for A2 as for A1, but the intermediate terms in the corresponding calculation (\ref{eq:sumA1}) are all exactly identical.  (The exact time symmetry ensures that the intermediate terms would continue to be identical even if one did not coarse-grain the fine-grained paths of (\ref{eq:fpi3}) into the four discrete paths of (\ref{eq:sumA1}) and (\ref{eq:sumA2}).)  From this identical mathematics under a well-defined spacetime transformation, FISH implies that any realistic description of one experiment will have a dual description in the other experiment.  It follows that any ``intuitive'' explanation of one will necessarily carry over to the other.  (As we are about to see, however, there is no guarantee the dual explanation will still be considered ``intuitive''.) 

For these two action-dual experiments, FISH simply requires two ontologies that map onto each other under the same spacetime transformation that maps path-integral trajectories in A1 to the equivalent set of trajectories in A2.  (A time-reversal, in this case.)  Note that the boundaries must also be consistent, such that any interpretation of an $A$-outcome in A1 maps only to the $C$-outcome in A2.  The central point is this: If some particular ontology insists that our A2-intuition is correct, and that nothing passes through $E$ for a $C$-detection in A2, then FISH can be used to construct a corresponding ontology for an $A$-detection in A1.   Time-reversing the intuitive picture of A2, the action-dual interpretation of A1 implies that nothing passes through $E$ on the way from $C$ to $A$. 

When interpreting A1 in this manner, note that one does not know which ontology to use until the experimental outcome has been registered.  If one detected a $B$-outcome instead, the action dual experiment would not be A2, but rather a version of A2 where the photon was sent into the other input channel.  This in turn would lead to a different ontological conclusion.  Therefore, respecting the action duality necessarily requires a model with ``retrodiction'', where the experimental outcome enables us to update our best-possible description of prior events.  For example, one might look at an $A$-outcome in A1 and mentally trace this result backward through the final interferometer, concluding (in hindsight) that the photon had actually bypassed $E$ entirely.

These dual experiments make it evident that any ontology based directly upon the standard QM wavefunction will be fundamentally incompatible with FISH.  In A1 a non-zero wavefunction passes through $E$, while in A2 the corresponding wavefunction at $E$ is exactly zero.  Not only does this explicitly violate FISH, but the retrodictive solution from the previous paragraphs appears to be unavailable.  Even after an $A$-outcome is registered in A1, there is no mechanism in standard QM to retrodict a more precise description of what has just happened (as there are generally no unknown parameters to update).  

Despite the failure of standard QM to yield a realistic ontology that respects the action duality, there are non-standard models where the necessary retrodiction is allowed; both the time-symmetric quantum mechanics (TSQM) of Aharonov \cite{Aharonov1,Aharonov2} as well as the de Broglie-Bohm (dBB) approach \cite{dBB}.  The geometry of this double-interferometer has previously been analyzed in this very context \cite{Hardy,Miller} where one finds that the full (retrodicted) ontology does respect the action duality in TSQM, and the particle ontology respects the action duality in dBB.  (Whether the quantum potential in dBB respects this same duality is a trickier question, one that will not be explored here.)  In searching for action-dual models of these experiments, one might also consider ontologies where something ``real'' passes through $E$ for both A1 {\it and} A2; in this case it is possible that no retrodiction would be necessary.

Although a thorough evaluation of these and other models are beyond the scope of this paper, it is worth addressing the larger question of how different models can yield identical quantum mechanical predictions (of experimental outcomes) and yet still vary in how well they conform to FISH.  After all, one might note that (\ref{eq:fpi}) is provably equivalent to  Schr\"odinger equation evolution followed by the Born rule applied to measurement outcomes \cite{Feynman, FeynmanHibbs, Schulman}; how then could interpretations of equivalent mathematical frameworks have any differences at all?  

The answer to this question lies in the intermediate steps of the calculations.  Examining (\ref{eq:fpi}) in detail, one sees that it is the $x$-integral that corresponds to the Schr\"odinger evolution, and the $y$-integral that corresponds to the Born rule.  The Schr\"odinger picture, therefore, insists that the integrals in (\ref{eq:fpi}) must be performed in this precise order, with an implied significance for the intermediate step
\begin{equation}
\label{eq:schr}
\psi(y;t_f)=\int \left(\sum_{x,t_i\to y,t_f} e^{iS/\hbar}\right)  \psi_i(x) \, dx.
\end{equation}
It is this ``evolved-but-uncollapsed" wavefunction $\psi(y;t_f)$ that the Schr\"odinger picture is built upon, and $\psi(y;t_f)$ is the most obvious object to assign an ontological status in this picture.  However, the path-integral equation (\ref{eq:fpi}) does not imply any importance for this intermediate step; for example, it would be perfectly acceptable to perform the two integrations in the opposite order.  In that case, a map back to a Schr\"odinger-like picture would appear as evolving $\psi^*_f(y)$ via the time-reversed Schr\"odinger equation, back to a different intermediate step $\psi^*(x;t_i)$, followed by the integral of $\psi^*(x;t_i)\psi_i(x)\,dx$.  

With this perspective, it should be clear that a literal interpretation of an intermediate $\psi$ does not conform to FISH because a time-reversal is equivalent to changing the order of integration in (\ref{eq:fpi}).  If one attempts to physically interpret the result of the first integral performed, then one will get a different answer upon time-reversal.  The path-integral equation (\ref{eq:fpi}), on the other hand, implies that the order of integration has no ontological significance, and is therefore fully symmetric upon time-reversal. 

From the example in this section, it should be clear that FISH is a principle that significantly constrains realistic interpretations of quantum phenomena.  The above analysis is also an example of how FISH can be used in a constructive manner: from a straightforward interpretation of A2, the action duality directly constructs a retrodictive interpretation of A1.  (Of course, any interpretation constructed via FISH is only as valid as the dual interpretation used in its construction.)  In the next section, a different pair of action-dual experiments will demonstrate that the action duality can be applied beyond mere time-symmetry.

\section{Action Duality for Entangled Systems}

Sinha and Sorkin have previously used the sum-over-histories framework to address quantum nonlocality by explicitly calculating Bell inequality violations in a two-photon experiment \cite{Sorkin}.  While the calculation proceeds by virtue of (\ref{eq:fpi3}), summing over entirely local paths, the ``nonlocality'' enters in the form of the action $S$ itself, which is an integral over spacetime and therefore implicitly nonlocal.

A simplified version of the Sinha/Sorkin experimental geometry is shown in Figure 3 (Experiment B1).  The main simplification is that the source $Z$ is here constrained to emit two photons in a back-to-back geometry (say, by annihilating an electron and a positron, or via post-selection of the source's momentum).  This was the effective result of another post-selection used by Sinha and Sorkin, but simply asserting it from the outset makes this geometry simpler.  Given this source constraint, one photon will be detected at either $A$ or $B$, and the other photon will be detected at either $C$ or $D$.  Two phase-delay elements are also shown, one located near the $A/B$ detectors, and the other near the $C/D$ detectors.  Note that the source produces ``which-way'' entanglement only; there is no constraint on the relative phase or polarization of the two photons.

\begin{figure}[htb]
\centerline{\includegraphics[width=.4\textwidth]{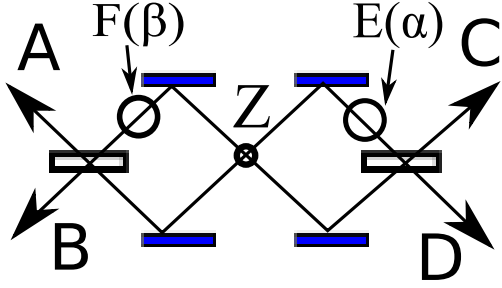}}
\caption{Experiment B1: A two-photon source $Z$ (constrained to emit in opposite directions only), sends one photon to the detectors at $A/B$ and the other to the detectors at $C/D$.  The circles represent phase delay elements $E$ and $F$ set to phase delay angles $\alpha$ and $\beta$ respectively.}
\label{Figure:b1}
\end{figure}

We will now reprise Sinha and Sorkin's ``sum over histories'' calculation \cite{Sorkin}.  Each joint probability only has two terms in the sum because there are only two coarse-grained scenarios to consider (due to the constraint on the source, exactly one photon will pass through a phase-delay element).  Also, because the action is just a spacetime integral over the entire region in question, the actions of the two photons are additive, and one can generate the two-photon action by adding the net phase delay of the two single-photon paths.  Applying (\ref{eq:fpi3}) then yields
\begin{align}
\label{eq:sumB}
P(A,C) & = \left| e^{i(\beta+\pi/2)}+e^{i(\alpha+\pi/2)}  \right|^2 \\
P(A,D) &= \left| e^{i(\beta+\pi)}+e^{i(\alpha)}  \right|^2 \\
P(B,C) & = \left| e^{i(\beta)}+e^{i(\alpha+\pi)}  \right|^2 \\
P(B,D) &= \left| e^{i(\beta+\pi/2)}+e^{i(\alpha+\pi/2)}  \right|^2.
\end{align}

Normalizing these results, the joint probability that both photons will emerge on the top or both on the bottom is $P(A,C)+P(B,D)=cos^2(\alpha/2-\beta/2)$.  The other joint probabilities $P(A,D)+P(B,C)$ sum to $sin^2(\alpha/2-\beta/2)$.  These probabilities violate the Bell inequality; the correlations therefore cannot be explained by any local hidden variable model if the hidden variables in common between the two photons are uncorrelated with the phase angle settings $\alpha$ and $\beta$.  Using a realistic local model to explain such experiments is therefore thought to be among the most difficult tasks in quantum foundations.

It turns out that Experiment B1 has an action dual in Experiment B2 (see Figure 4).  Here everything is the same as B1 except that the source $Z$ has been removed; the input is now a single photon at $C$ (or instead $D$, if one chose).  It is perhaps less obvious that B1 and B2 are related through an action duality; in B2 a single photon (traveling twice the distance) accumulates the same action/phase as two photons accumulate together in B1.  Technically, B2 is formed by time-reversing the right-hand portion of B1 (the portion physically to the right of $Z$ in Figure 3), pivoting around the fixed-point $Z$.  As the classical action is just an integral over the spacetime region spanned by the experiment, the time-symmetry of the classical Lagrangian density implies an action duality between B1 and B2.  After such a transformation, the source $Z$ simply becomes a crossing point, and the previous constraint that the B1 photons must be emitted in opposite directions is dual to the fact that (without a mirror or beamsplitter) there is no probability that the single photon will change course at the central crossing point in B2.  

\begin{figure}[htb]
\centerline{\includegraphics[width=.4\textwidth]{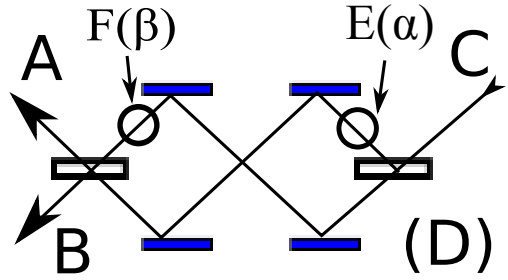}}
\caption{Experiment B2: an interferometer for a single photon input at $C$.   The circles represent phase delay elements $E$ and $F$ set to a phase delay angles $\alpha$ and $\beta$ respectively; the photon will pass through exactly one of these.}
\label{Figure:b2}
\end{figure}

As the classical action is the same for B1 and B2 (given the corresponding inputs/outputs), the joint probabilities must be identical as well; indeed every single term in the intermediate calculation (\ref{eq:sumB}) is also identical.  The joint probabilities in B2 continue to violate the Bell inequality, of course, but this is no longer of particular concern because the premise behind the derivation of the inequality seems obviously false in this case.  One would hardly expect to demand that any hidden variables at the central crossing-point in B2 to be independent of $\alpha$.

Despite our different instinctive judgements of the ``spookiness'' of the correlations in B1 and B2, the path-integral mathematics contains no differences at all.  The constructive benefit of applying FISH to the action duality is that one can take any realistic model of B2 and use the same spacetime transformation of the action to generate a corresponding realistic model of B1.  So, for example, if one assigns some realistic state to the halfway point in B2 that encodes the experimental setting of $E(\alpha)$, then there is a dual explanation of B1 where there are realistic states of the photons at $Z$ that also encode $\alpha$.  In B1, $E$ is now in the future of $Z$, which means any such model would seem to have retrocausal elements; the action duality has mapped an instinctively-acceptable model to an instinctively-wrong model.  And yet, the mathematics is identical in these two cases, perhaps implying that it is our {\it instincts} that are wrong.  Further discussion of the retrocausal implications of the action duality can be found in Evans, Price, and Wharton. \cite{EPW}

Apart from these retrocausal issues, there are two additional objections to the above conclusions that may be raised at this point.  The first is that in standard QM, the state space is not the same between one-particle and two-particle experiments, and therefore one could not expect to map one of these onto the other under {\it any} duality.  The second is the concern that the particular example in this section (along with the previously-published example \cite{EPW}) are ``special cases'' that could not be used to address Bell inequality violations in more general geometries.  Both of these concerns are addressed in the following section.

\section{Generalizations}

In this section, we show that the probabilities of the outcomes of measurements are the same for any two experiments related by the action duality defined in section II. There are two cases to consider: single quantum systems (as in section III) and entangled systems (as in section IV).  Note that this section's use of Dirac ket-notation and QM operators is mostly for convenience. Every matrix element has an equivalent path-integral form \cite{Schulman} to which the action duality can be directly applied. 

\subsection{Single quantum systems}

In the following, the time-reversal operator $\Theta = RK$ where $K$ is complex conjugation and $R$ is a unitary operator which depends on the representation.  In version I of the experiment involving a single quantum system exhibiting the duality (e.g. Experiment A1, above), if the initial state at some arbitrary time $t_i$ is $|\psi_i;t_i \rangle$ the probability that outcome $|\psi_f;t_i+\Delta t \rangle$ is measured at time $t_f=t_i+\Delta t$ is
\begin{equation}
P_I(\psi_i,\psi_f)=|\langle \psi_f;t_i+\Delta t| U(t_i+\Delta t,t_i)|\psi_i;t_i \rangle|^2 
\end{equation}
where $U(t_i+\Delta t,t_i)$ is the time-evolution operator which depends on the Hamiltonian $H_I(t)$. Experiment A1 consists of a sequence of time independent Hamiltonians $H_n$, each acting over time $\delta t_n$ so that
\begin{equation}
\label{eq:Uproduct}
U(t_i+\Delta t,t_i)=U_1(\delta t_1)U_2(\delta t_2) \ldots U_N(\delta t_N)
\end{equation}
where $U_n(\delta t_n)=\exp(-iH_n\delta t_n)$ and $\Delta t=\sum_{n=1}^{N} \delta t_n$. Since the Hamiltonians are time-independent the choice of $t_i$ is arbitrary both in version I and the dual experiment but it is retained in the following for clarity.

In the dual experiment (e.g. Experiment A2, above), the quantum system is prepared in state $|\tilde{\psi}_f;-t_i-\Delta t \rangle=\Theta|\psi_f;t_i + \Delta t \rangle$ and we require the probability \begin{equation}
P_{II}(\tilde{\psi}_i,\tilde{\psi}_f)=|\langle \tilde{\psi}_i;-t_i| U'(-t_i,-t_i-\Delta t)|\tilde{\psi}_f;-t_i-\Delta t \rangle|^2
\end{equation}
that the outcome $|\tilde{\psi}_i;-t_i \rangle=\Theta|\psi_i;t_i \rangle$ will be obtained after the evolution $U'(-t_i,-t_i-\Delta t)$ which is determined by the Hamiltonian $H_{II}$ in the dual experiment. From a well-known identity \cite{Messiah}
\begin{align}
\label{eq:identity}
\langle \psi_f;t_i+\Delta t| U(t_i+\Delta t,t_i)|\psi_i;t_i \rangle & =\langle \tilde{\psi}_i;-t_i| \Theta U^{\dagger}(t_i+\Delta t,t_i) \Theta^{-1} |\tilde{\psi}_f;-t_i-\Delta t \rangle \\
& = \langle \tilde{\psi}_i;-t_i| U_N(\delta t_N) \ldots U_2(\delta t_2) U_(\delta t_1)  |\tilde{\psi}_f;-t_i-\Delta t \rangle
\end{align}
where the second line follows from Eq.~(\ref{eq:Uproduct}) because $\Theta U^{\dagger}(\delta t)\Theta^{-1}=U(\delta t)$ \cite{Messiah}. Therefore the probabilities $P_I(\psi_i,\psi_f)$ and $P_{II}(\tilde{\psi}_i,\tilde{\psi}_f)$ are equal provided the Hamiltonian $H_{II}$ for the dual experiment is constructed so that it is the sequence of Hamiltonians in $H_I$ applied in reverse time-order. (The case of time-dependent Hamiltonians can be dealt with using the formalism in Ref.~\cite{Gross} and leads to the requirement that $H_{II}(t-t_f)=H_I(t_f-t)$.)

\subsection{Entangled systems}

When considering action dualities, it is crucial to maintain a spacetime perspective.  With that in mind, the framework for this subsection will be a comparison between the two generic experiments shown in Figure 5.  In the first experiment (Figure 5a) is a bipartite entangled system subject to some Hamiltonian $H=H_R+H_L$; here $H_R$ is assumed to only affect the subsystem on the right hand side (rhs), and $H_L$ only affects the subsystem on the left hand side (lhs).  We only consider cases where two maximally-entangled subsystems have the same Hilbert space dimension ($d$).  An initial preparation at time $t=t_i$ fixes the initial condition to be an entangled state $\ket{\Psi;t_i}$, while final measurements on the two subsystems yield the results $\ket{\alpha;t_L}$ on the lhs and $\ket{\beta;t_R}$ on the rhs.  The joint probabilities of the outcomes $P(\alpha,\beta)$ may or may not violate the Bell Inequality.

\begin{figure}[htb]
\centerline{\includegraphics[width=.4\textwidth]{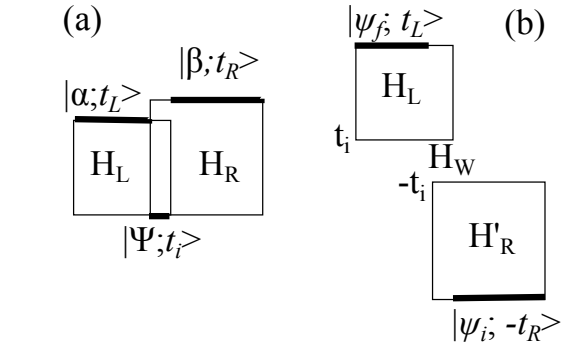}}
\caption{Two generic, action-dual experiments; time runs vertically, one spatial dimension is horizontal, the other spatial dimensions are suppressed.  (a) An initially-entangled bipartite system $\ket{\Phi}$ is spatially separated, with each subsystem measured at a later time. (b) A single system is subject to a unitary evolution before being measured.  See text for further details.}
\label{Figure:5ab}
\end{figure}

There is no entanglement in Figure 5b; this is simply a single system with Hilbert space dimensionality $d$ subject to three sequential Hamiltonians: $H'_R$, $H_W$, and lastly $H_L$.  The interaction times are chosen such that $H_L$ is acting for the same duration in both Figure 5a and 5b; this permits the two $H_L's$ to be identical.  $H_R$ and $H'_R$ also act for the same duration. In moving from Figure 5a to 5b, we will make the identification
\begin{equation}
\label{eq:trs}
\Theta \ket{\beta;t_R}=\ket{\psi_i;-t_R}.
\end{equation}

The role of $H_W$ in Figure 5b is analogous to the role of the initial entangled state $\ket{\Psi}$ in 5a, in that the joint probabilities $P(\psi_i,\psi_f)$ are dependent upon a given $H_W$ in the same way that $P(\alpha,\beta)$ is dependent upon a given $\ket{\Psi}$.  Indeed, a preparation of an entangled state can be viewed as effectively a constraint that relates measurement outcomes from the two subsystems, just as the Hamiltonian $H_W$ is a constraint that relates the output of the large square and the input to the small square in Figure 5b.  

Recall that the action duality applies when there are two experiments with a spacetime mapping between the two classical Lagrangian densities; given identical boundaries, this leads to an identical classical action, from which it follows from (\ref{eq:fpi}) that the joint probabilities must be identical.  The question then is whether the two experiments in Figure 5 generally have such a mapping.  It should be clear that there is no such map if there is any interaction between the two subsystems in Figure 5a (after the initial measurement).   This is because an interaction term in the Lagrangian would have no analog in Figure 5b, where the spacetime region comprising the action has been split into two disjoint regions.  Nevertheless, it is typical for an entanglement experiment to immediately separate the two subsystems after the initial measurement, and in this case recovering an action duality seems generally plausible.  The smaller square in both 5a and 5b would comprise the Lagrangian corresponding to $H_L$, one dimension-$d$ subsystem, and the interaction between them; the larger square would comprise $H_R$, one subsystem, and that other interaction.

The remaining complication is the small overlap between the two squares in Figure 5a; this overlap implies some spatial extent of the initial entangled state, and it is possible that this region could break the action duality.  For example, even if there was no direct interaction between the two subsystems after the initial measurement, if there were indirect interactions (say, via some background field in this overlap region) it would be impossible to replicate such a Lagrangian density in Figure 5b.  That said, if even the indirect interactions were negligible, then this overlap is irrelevant, and the action duality could be recovered. Experiments to investigate correlations involving entangled particles, including the example in the previous section, are designed to avoid these caveats; not only do the two photons in Figure 3 have no mutual interactions (direct or indirect), but it also turns out that the unitary evolution corresponding to $H_W$ is simply the identity.   

It remains to calculate the unitary operator corresponding to $H_W$; first consider Figure 5a.  For subsystems with Hilbert spaces of the same dimensionality $d$, the most general bipartite entangled state at time $t=t_i$ can always be written in the Schmidt decomposition
\begin{equation}
|\Psi;t_i \rangle = \sum_n c_n |u_n;t_i \rangle |v_n;t_i\rangle
\end{equation}
where $|u_n;t_i \rangle$ and $|v_n;t_i \rangle$ are some basis for each subsystem.  The first basis applies to the subsystem sent to the lhs detector, measured at time $t_L$; the second basis applies to the subsystem sent to the rhs detector, measured at time $t_R$.  We assume that the observable measured at each detector is given (although the outcome, of course, is not). 

The time evolution of the lhs subsystem (from $t\!=\!t_i$ to $t\!=\!t_L$) results from the unitary time-evolution operator $U_L(t_L,t_i)$ that corresponds to $H_L$ in Figure 5a.  This evolution is followed by a projective measurement resulting in the state $|\alpha;t_L \rangle$.  Similarly, the unitary time-evolution operator on the rhs is $U_R(t_R,t_i)$ and the measured state on the rhs is $| \beta;t_R \rangle$. The joint probability of outcomes $| \alpha;t_L \rangle$ and $|\beta;t_R \rangle$ is 
\begin{equation}
\label{eq:prob}
P_{\Psi}(\alpha, \beta) = \left |\sum_n c_n \langle \alpha;t_L|U_L(t_L,t_i)|u_n;t_i \rangle \langle \beta;t_R|U_R(t_R,t_i)|v_n;t_i \rangle \right |^2 .
\end{equation}

From the identity used to obtain Eq.~(\ref{eq:identity})
\begin{equation}
 \langle \beta;t_R|U_R(t_R,t_i)|v_n;t_i \rangle = \langle \tilde{v}_n;-t_i | U'_R(-t_i, -t_R) |  \tilde{\beta};-t_R \rangle,
 \end{equation}
where $U'_R(-t_i, -t_R)$ is the time-evolution operator generated from applying the Hamiltonian on the rhs in the reverse time-sequence.  But the final ket in this expression has already been identified as $\ket{\psi_i;-t_R}$ in (\ref{eq:trs}).  Together with the above-discussed assignment $\ket{\alpha;t_f}=\ket{\psi_f;t_f}$, this means that (\ref{eq:prob}) can be rewritten as
\begin{align}
\label{eq:key}
P_{\Psi}(\alpha, \beta)  & =   \left |  \sum_n c_n  \langle \psi_f;t_L|U_L(t_L,t_i)|u_n;t_i \rangle  \langle \tilde{v}_n;-t_i| U'_R(-t_i, -t_R)  |\psi_i; -t_R \rangle \right |^2 \nonumber \\
& =  \left | \langle \psi_f;t_L|U_L(t_L,t_i)\, W(t_i,-t_i) \, U'_R(-t_i, -t_R)  |\psi_i; -t_R \rangle \right |^2 ,
\end{align}
where
\begin{equation}
W(t_i,-t_i)= \sum_n c_n |u_n;t_i \rangle  \langle \tilde{v}_n;-t_i|.
\end{equation}
If $\ket{\Psi}$ is maximally entangled (as applies in section V), $\sqrt{d}\,W(t_i,-t_i)$ is a unitary operator. As mentioned in the previous subsection, the choice of $t_i$ is arbitrary.

The main result is that if $\sqrt{d}\,W(t_i,-t_i)$ is the unitary operator that corresponds to $H_W$ in Figure 5b, then
\begin{equation}
P_{\Psi}(\alpha,\beta) = \frac{1}{d} \, P_W[\psi_f|\psi_i].
\end{equation}

Here $P_W[\psi_f|\psi_i]$ is the usual conditional probability that outcome $\psi_f$ will be measured at $t_L$ from a quantum system prepared in the state $\psi_i$ at $-t_R$ and subject to the unitary transformations $U'_R$, $\sqrt{d}\,W$, and then $U'_L$ (as shown in 5b).  The extra factor of $d$ is irrelevant because of the usual normalization procedure used on joint probabilities. Hence we have shown the equivalence of the probabilities calculated from Figs.~5a and 5b.

The equivalence of the probabilities $P_W[\psi_f|\psi_i]$ for the single system in Fig.~5b and $P_{\Psi}(\alpha, \beta)$ for the entangled system in Fig.~5b is quite similar to a general isomorphism recently demonstrated by Leifer \cite{Leifer} (and in turn based on an isomorphism noted by Jamiolkowski \cite{Jam}).  Although a different implementation makes those results appear somewhat different from the present construction, forthcoming work \cite{LnS} implies that the above results are consistent with Leifer's if one interprets the transpose as a time-reversal operation.

\section{Conclusion}

FISH is a constraint on realist models of quantum phenomena. It proposes that there exists an \emph{ontological} connection between certain pairs of experiments that are normally taken to involve only an ``accidental'' symmetry at the level of observable correlations. The motivation for exploring the consequences of this hypothesis is the intuition that the symmetries of the path integral might be a clue to the symmetries of the underlying ontology of quantum phenomena, in a deeper sense than is possible in realist models that do not respect FISH. We emphasize that this motivation is far from providing an argument that FISH is true.  But it does, in our view, provide a strong reason to entertain the possibility that FISH \emph{might} be true -- in other words, a reason to explore FISH-friendly models of quantum phenomena, even if they seem initially to be counterintuitive in other respects.

As a specific example of how FISH can offer a different perspective than standard QM, consider the treatment of particle number in standard quantum theory.  Our intuition tells us that there is a crucial difference between a system with two particles at one time (such as Experiment B1) and a system with only a single particle at any time (such as Experiment B2).  Standard QM bears out this intuition by representing these two experiments via quite different mathematical structures.  Two simultaneous particles reside in a multidimensional configuration space (say, $\psi(x_1,x_2)$), but there is no analogous structure in standard QM for two particles at different times (or one particle at two points on its worldline, say $\phi(t_1,t_2)$).  Experiment B2 is therefore described using a space of smaller dimensionality, say $\psi(x_1)$.  

But has our intuition led us astray?  Hardy has pointed out that there is no clear way to choose the appropriate dimensionality of the Hilbert space in curved-spacetime systems where the causal structure is not known \textit{a priori} \cite{Hardy2}, and FISH now adds to the problem.  Given the existence of an spacetime representation of one of two action-dual experiments (Experiment B2), denying the possibility of a similar representation for Experiment B1 explicitly violates FISH.  Insofar as it is the particle-counting intuition that leads to the broken symmetry, we might learn something by discarding this intuition and trusting the symmetries of the path integral.  And in that mathematics, there is no fundamental difference between Experiments B1 and B2.  Indeed, it was Feynman's willingness to view particle number as a subjective parsing of an entire spacetime history that helped lead to his version of quantum electrodynamics, as in his famous example: 

\begin{quote} Following the charge rather than the particles corresponds to considering this continuous world line as a whole rather than breaking it up into its pieces.  It is as though a bombardier flying low over a road suddenly sees three roads and it is only when two of them come together and disappear again than he realizes that he has simply passed over a long switchback in a single road.\cite{Feynman2} \end{quote}

We may distinguish two means of incorporating FISH into current debates. (There may well be other ways of doing so, but these will do for our purposes.) The first approach begins with the standpoint of the quantum realist, who accepts, at least provisionally, that there is a real ontology underlying quantum phenomena, and is looking for clues about the nature of that reality. The second begins with that of the quantum Òanti-realistÓ, who believes that the realist program is mistaken, and is looking for Òno-go argumentsÓ, to use against it.

For anti-realists, FISH is a significant challenge. It provides a technique for constructing realist models -- and a \emph{prima facie} justification, based on symmetry considerations, for taking them seriously -- in places where the anti-realist might have hoped to have ruled them out. (It allows local realist explanations of Bell correlations, for example.) Its importance is therefore that it significantly weakens the case for anti-realism, by weakening the no-go theorems. To meet this challenge -- to restore those arguments to their prior strength -- the anti-realist needs an argument \emph{against} FISH; in other words, a reason for thinking that quantum ontology could not respect the symmetries of the path-integral.

For realists, things are a bit different. From the realist perspective, FISH is a proposal about the nature of any ``true'' ontology, underlying quantum phenomena. Here, too, it has a challenging aspect, in that its recommendations may seem in conflict with our intuitive causal picture.  But it is also potentially an opportunity, in the sense that it leads us to consider possibilities that we would otherwise ignore \emph{because} they are counterintuitive.  Some intuitions deserve to be challenged, after all, and symmetry is often a useful guide to what needs to go. 

Some realists may be tempted to join forces with anti-realists, in seeking a decisive argument against FISH. We cannot exclude the possibility that some such argument might be found. In its absence, however, neither realists nor anti-realists can afford to ignore the possibility that FISH may embody a deep truth about the quantum world, and that the implications of Feynman's integral for quantum reality may lie in its symmetries.

\section*{Acknowledgements}  The authors are grateful to the useful comments provided by E. Cavalcanti, J. Finkelstein, M. Leifer, and especially both O. Maroney and R. Sutherland.  KW gratefully acknowledges that much of this joint research was conducted during his tenure of a Visiting Fellowship at the Sydney Centre for the Foundations of Science, of the University of Sydney.


\bibliographystyle{mdpi}
\makeatletter
\renewcommand\@biblabel[1]{#1. }
\makeatother

\end{document}